\documentclass[twocolumn,aps,prb,showpacs]{revtex4}

\usepackage{graphicx}
\usepackage{color}
\usepackage{amssymb}

\def\bra#1{\langle #1|}
\def\ket#1{|#1\rangle}
\def\braket#1#2{\langle #1|#2\rangle}
\def\vp{\varphi}
\def\vr{\varrho}
\def\omegaeff{\omega_{{\rm eff}}}
\def\Tr{\mathop{\rm Tr}\nolimits}
\def\op#1{\hat{#1}}

\def\ds{\displaystyle}
\def\d{{\mathrm d}}

\def\unit#1{\ \mathrm{#1}}
\def\sgn#1{\mathop{\rm sgn}}

\definecolor{orange}{rgb}{0.7,.35,0}

\begin{document}

\preprint{Version 26.04.2002}

\title{Magnetoresistance Calculations for a Two-Dimensional Electron Gas
with Unilateral Short--Period Strong Modulation}

\author{Karel V\'yborn\'y}
\affiliation{I. Institute of Theoretical Physics, University
of Hamburg, Jungiusstr. 9, D--20355 Hamburg, Germany}
\email{vybornyk@fzu.cz}

\author{Ludv\'\i k Smr\v cka}
\affiliation{Institute of Physics, Academy of Sciences of the Czech Republic,
Cukrovarnick\'a 10, CZ--16253 Praha, Czech Republic}

\author{Rainer A. Deutschmann}
\affiliation{Walter Schottky Institut, Technische Universit\"at 
M\"unchen, D--85748 Garching, Germany}
\altaffiliation{now with McKinsey\&Company}

\begin{abstract}

\vspace{5mm} 
  The linear response theory is used to describe magnetoresistance
  oscillations of short--period unilateral superlattices with strong
  modulation (or alternatively arrays of coupled quantum wires). The
  semiclassical description of this system fails for strong magnetic
  fields (magnetic breakdown) and we employ a simple
  fully--quantum--mechanical tight--binding model in conjunction with
  Kubo's formula instead. The resulting magnetoresistance data nicely
  compare to the experiments while the model opens good intuitive
  insight into the effects taking place in the system.
\end{abstract}


\maketitle

\section{Introduction}

Transport properties of two--dimensional electron systems (2DES) with
unidirectional periodical modulation have been studied for more than
ten years. Lot of attention has been paid to the case of weak
modulation by electric and magnetic fields. Such a system was first
prepared by means of holographic techniques by Weiss {\it et al.}
\cite{weiss:01:1989} and showed the commensurability oscillations in
magnetoresistance for low magnetic fields and Shubnikov--de Haas (SdH)
oscillations for high magnetic fields. Periodicity of the former ones
can be well understood even in a semiclassical (SC) concept
considering the drift of the cyclotron orbit centre in crossed
electric and magnetic fields. This approach can also give some
quantitative predictions for the magnetoresistance
\cite{beenakker:04:1989}. An alternative formulation of the SC
approach \cite{streda:06:1990} relying on the breakdown probability
(tunnelling between two closed SC orbits) oscillations is also
possible.  Gerhardts {\it et al.}  \cite{gerhardts:03:1989}
diagonalised the quantum--mechanical (QM) Hamiltonian (and employed
the Kubo formula to compute conductivity) finding the oscillating
width of Landau bands (Landau levels broadened by the weak modulation
into narrow cosine--like bands) to be the basic cause of the effect
within the QM picture.  QM approach was also applied by Vasilopoulos
{\it et al.}  \cite{vasilo:11:1989}.

The SdH oscillations follow naturally from the QM concept owing to the
quantization of free electron motion in magnetic fields (Landau levels,
LLs). This quantization has to be {\em ad hoc} assumed in the SC
picture but once this is accepted, the SC theory provides a sufficient
description in this case.

Experiments on gated structures manufactured by lithographic
techniques performed by Beton {\em et al.} \cite{beton:11:1990}
allowed for investigating effects of stronger modulation. Compared to
the previously mentioned experiments, the major effect was the
quenching of the commensurability oscillations.  The SC approach
\cite{beenakker:04:1989} was applicable again \cite{beton:11:1990},
although quantum calculations \cite{shi:05:1996} were in better
agreement with experiments.  Later, the SC theories have been extended
in order to account for anisotropic scattering \cite{menne:01:1998}
and for modulation by magnetic field
\cite{zwerschke:00:1998,zwerschke:08:1998} (extending the older works
\cite{beenakker:04:1989,muller:01:1992}).  For QM approach to
modulation by magnetic field see e.g. \cite{peeters:01:1993}. All
together, the SC approach proved itself to be suitable in these cases.

More recently, samples with modulation potential due to MBE--grown
structure were prepared (using cleaved edge overgrowth technique) by
Deutschmann {\em et al.}\cite{deutschmann:02:2000} In contrast to all
former experiments, the modulation period $d=15\unit{nm}$ was shorter
by almost one order of magnitude. Owing to this fact and also due to
strong modulation (Fermi energy $E_F\simeq 4|t|$, where $t$ is the
hopping integral between ground states in two neighbouring wells of
the potential modulation), the lowest modulation miniband is well
separated from higher minibands (condition for this is $d\lesssim
\sqrt{3h^2/(2|t|m)}$).  The magnetoresistance oscillations measured
when the Fermi level lies between the modulation bands can be
explained by no semiclassical model unless tunnelling between open
trajectories is assumed (the breakdown formalism\cite{streda:06:1990}
mentioned above is necessary at this place). It is thus appropriate
to revert to a quantum mechanical description.  Moreover, the miniband
structure is simple now and allows thus for a good insight into the
physics both on the SC and quantum--mechanical level.

\subsection{The System and Outline}

\begin{figure}
\begin{center}
\includegraphics[scale=0.4]{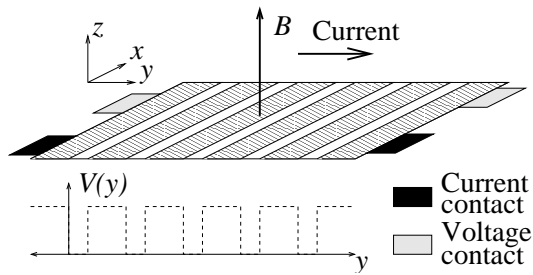} 
\end{center}
\caption{ 2DES with lateral periodical modulation and 
the effectively two--point-contact geometry. }
\label{fig1_1}
\end{figure}

In this paper, we refer to experiments carried out on GaAlAs/GaAs
structures first reported in \cite{deutschmann:02:2000}, see also more
detailed description in \cite{deutschmann:02:2001,deutschmann:2001}.
These are superlattices with strong unilateral short--period
(electric--field) modulation. The substantial difference to previous
studies (see the Introduction) is the shortness of the modulation
period: only the lowest modulation band (and not many of them) is
occupied under those circumstances which makes the usage of quantum
mechanics unevitable. Moreover, absence of the higher modulation bands
makes the model very transparent.  We concentrate on magnetoresistance
measurements at different concentrations of electrons (which could be
varied by a gate voltage over the range $0.5\div 5.0\times
10^{11}\unit{cm^{-2}}$). The system is sketched in Fig. \ref{fig1_1}.

The band structure in zero magnetic field calculated within the Kronig--Penney
model can be well approximated by
\begin{equation}\label{eq1_1}
  E(k_x,k_y)=\frac{\hbar^2k_x^2}{2m}-2|t|\cos k_yd
\end{equation}
with $4|t|\approx 3.8\unit{meV}$  and $m$ equal to the effective
mass of electrons in GaAs. The next modulation band
is well above the Fermi level $E_F$ for all accessible concentrations.

If the Fermi level lies near the bottom of the band ($-2|t|<E_F\ll
2|t|$), the system resembles a free 2DES (i.e. paraboloidal spectrum
with modified effective mass in the modulation direction
$m_y=\hbar^2/2|t|d^2 \approx 2.7m$).  
If the Fermi level lies high above the
modulation band edge ($E_F\gg 2|t|$), the dispersion relation is like
the one of an array of 
separated one--dimensional wires ( i.e. parabolic across $k_x$ and
nearly constant along $k_y$ )  
while the deviation from $E=\hbar^2k_x^2/2m$ reflects the coupling of wires.

The structure of this paper is the following.  At the beginning we
will review the semiclassical approach as a favourite tool used to
describe magnetotransport experiments. We will  also compare the
zero--field density of states (DOS) with the $B\not =0$ DOS computed
in Sec. \ref{secII}. It will give us illustrative examples of a
situation when the SC theory is expected to be successful and of
another situation when it should fail. The condition of applicability
of the SC approach will be shown to be $\hbar\omegaeff\ll 2|t|$
whereas $\omegaeff=eB/m_{{\rm eff}}=eB/\sqrt{mm_y}$. 

In the second part (Sec. \ref{secII}) we will describe a
fully--quantum--mechanical  one--particle  model and
demonstrate that the gaps in the DOS emerging from this model coincide
with extrema in the measured magnetoresistance. Then we will employ
the linear response theory (section \ref{secIII}) in order to
calculate the magnetoresistance and we will compare it with the
experiments.

\subsection{Semiclassical Approach}

The physical quantity of central importance in transport theories is
the density of states $g(E)$ (DOS) at the Fermi level $E_F$. It is
known that its structure reflects features of the resistance (both as
a function of $B$, for instance) but the relation between these two
quantities is not simple. 

The SC theory attempts to explain the behaviour of electrons subject to
a magnetic field in terms of the zero--field Fermi surfaces.  We
demonstrate that the zero--field DOS plus an extra quantization
condition is a fairly good approximation to the realistic DOS
(computed by our model, from Section \ref{secII}) at low magnetic
fields ($\hbar\omegaeff\ll 2|t|$).  However, there is a drastic
difference between these two densities of states for high magnetic
fields indicating failure of the SC theory (see Fig. \ref{fig2_1}).

\begin{figure}[hb]
\begin{center}
\leavevmode
\hbox{\includegraphics[angle=-90, scale=0.3]{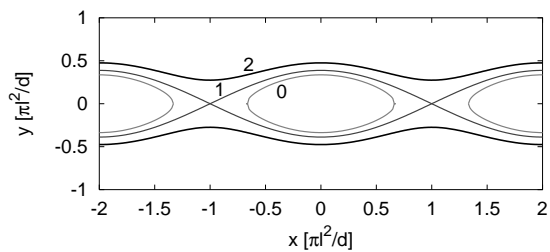}} 
\end{center}
\caption{ Real--space semiclassical trajectories of electron 
in magnetic field. 0 -- closed
($E_F<2|t|$), 1 -- critical ($E_F=2|t|$), 2 -- open
($E_F>2|t|$). The
closed trajectories are elongated in the direction parallel to the
wires by a constant factor $\sqrt{m_y/m}$ and due to non--parabolicity
of the cosine band (Eq.~\ref{eq1_1}).
}
\label{fig2_2}
\end{figure}

\begin{figure}
\begin{center}
\leavevmode
\hbox{\includegraphics[angle=-90, scale=0.3]{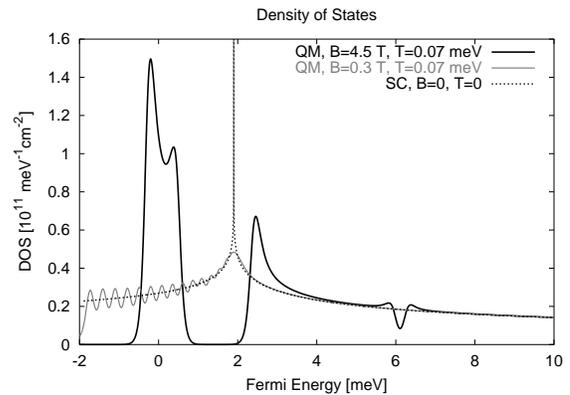}}
\end{center}
\caption{The density of states for $B=0$ (dotted) and thermally
broadened ($T\approx 1\unit{K}$) 
density of states for low field ($\hbar\omegaeff\ll 2|t|$ or 
$\alpha\ll 1$, solid grey line) and high field (solid black line).}
\label{fig2_1}
\end{figure}

Let us briefly review the SC approach suggested by Lifshitz and
Onsager (see e.g. \cite{ashcroft:1976:p232}).  We construct the
Fermi contour $E_F=E(k_x,k_y)$ for a given Fermi level. The statements
are that (1) the Fermi contour rotated by 90 degrees and scaled by
$\ell^2=\hbar/eB$ corresponds to the real--space trajectory of an
electron (see Fig. \ref{fig2_2}) and (2) if the contour is closed,
then it is allowed only if the magnetic flux passing through the area
enclosed by the real--space trajectory is an integer multiple of the
magnetic flux quantum $\Phi_0=h/e$.  In the QM picture, this
quantization condition corresponds to the situation when $E_F=\hbar\omegaeff
(n+\frac{1}{2})$ for some integer $n$, i.e. when the $n^{\rm th}$ LL
passes through the Fermi level. 

So as to be able to compare the SC and QM predictions let us now
examine the densities of states.  Note that by comparing the
zero--field DOS of our system (for $E\ll 2|t|$, i.e. in the region of
quantized orbits and 2D--like behaviour) with the zero--field DOS of a
free 2DES ($g(E)=2m_{\rm eff}/\pi\hbar^2$ including spin) we may
deduce the modulation--influenced effective mass $m_{\rm eff}$. This
in turn determines the quantization condition $E_F=\hbar\omegaeff
(n+\frac{1}{2})$ and thus all SC predictions can be made using the
zero--field DOS of the system only.  Based on the spectrum
(Eq. \ref{eq1_1}) we can compute the zero--field DOS analytically
\begin{equation}\label{ZeroFieldDOS}
\hskip-3mm
  g_0(E)=\left\{
\begin{array}{l}\displaystyle
 	\frac{4}{(2\pi)^2}\sqrt{\frac{2m}{\hbar^2 |t|d^2}}\cdot 
	\frac{1}{\sqrt{\xi}}
	K\big(1/\sqrt{\xi}\big)\,,\
         E>2|t|\
\\ \displaystyle
	\frac{4}{(2\pi)^2}
	\sqrt{\frac{2m}{\hbar^2 |t|d^2}}\cdot K(\sqrt{\xi})
 	\,,\ -2|t|<E<2|t|
	\end{array}\right.\hskip-5mm
\end{equation}
where
$$
\displaystyle \xi=\frac{1+E/2|t|}{2}
$$
including the factor of two for spin 
(the dotted line in Fig. \ref{fig2_1}). $K$ is the full elliptic function
$K(k)=F(\frac{\pi}{2},k)=\int_0^{\pi/2}(1-k^2\sin^2 \vp)^{-1/2}\d\vp$,
we recall\cite{ryzhik:1980:p904} that $K(0)=\pi/2$, $K(1)=\infty$. 

Let us focus on the weak--field case first ($\hbar\omegaeff\ll 2|t|$)
and discuss the influence of magnetic field $B$ on the continuous
spectrum (Eq.  \ref{eq1_1}), i.e. we try to estimate the DOS in
magnetic fields without any calculation.  On one hand, Landau levels
(LLs) appear for $E_F\ll 2|t|$ (2D--like paraboloid band structure,
modified effective mass $m_y$).  If the thermal energy is
comparable to the Landau level separation ($k_BT\approx \hbar\omega$) 
the DOS becomes oscillatory (leading to SdH oscillations) but
approaches the zero--field DOS (see also Fig. \ref{fig2_1}). In other
words, the oscillations missing in the zero--field DOS are exactly
reproduced by the SC quantization condition (closed trajectories,
$E_F<2|t|$) which is fulfilled just when $E_F$ lies in the middle
between two LLs.

On the other hand, narrow gaps open in the continuous spectrum for
$E_F\gg 2|t|$ due to the slight corrugation (by the cosine term) of
the almost 1D--like parabolic--trough band structure. Numerical
calculations (using model from Sec. \ref{secIII}) show that these gaps
are narrow enough to disappear due to thermal broadening and the
zero--field DOS matches the non--zero--field DOS perfectly. The SC
approach relying on non--quantized open trajectories ($E_F>2|t|$) and
predicting non--oscillatory magnetoresistance will therefore be
successful once again.

The picture is considerably different for strong magnetic fields
($\hbar\omegaeff\gtrsim 2|t|$) at which the cyclotron radius $R_c$
approaches $d$.  The numerically computed DOS shows no similarity to
the zero--field DOS (Fig. \ref{fig2_1} again). The non--zero--field
DOS rather resembles $g(E)\propto 1/\sqrt{E}$ of a single quantum wire
(suppression of tunnelling between two neighbouring wires when
classical cyclotron radii become comparable to $d$) with gaps both for
$E<2|t|$ and $E>2|t|$ owing to the tunnelling between the
wires. Therefore we expect oscillatory magnetoresistance for both
closed and open SC trajectories on contrary to the SC predictions
(magnetic breakdown in the SC theory).

The gaps occur at the boundary of the first magnetic Brillouin zone
(1MBZ, see the section \ref{secII}) and result from the periodicity of
the dispersion relation in $k_x$ which in turn reflects the invariance
of the QM Hamiltonian to magnetic translations
\cite{pfannkuche:06:1992} (this is what remains from the full
translational symmetry in the $x$--direction after switching on the
magnetic field). Magnetic breakdown can be included also in the SC
picture through tunnelling between two open or two closed
trajectories\cite{streda:06:1990} (see Fig. \ref{fig2_2}).  This is an
{\it ad hoc} assumption though.

The failure of the SC approach for strong magnetic fields is apparent
also in another context. Once the Fermi level is set and the type of
trajectory is determined, the same behaviour (either 2DES--like SdH
oscillations or 1DES--like no oscillations) is predicted for all
magnetic fields. However, it is clear even on the SC level that if the
cyclotron radius $R_c=k_F\ell^2$ ($k_F$ is the Fermi wavevector,
$E=\hbar^2 k_F/2m$) becomes comparable to $d$, tunnelling between
the wires is suppressed and the system ought to switch to a quasi--1D
regime.

The QM approach is suitable for both low and strong magnetic fields.
Moreover, it opens up the way to quantitative calculations of the
magnetoresistance and thus to data directly comparable to the
experiments.

\section{Model}
\label{sec2}\label{secII}

Having chosen the Landau gauge $\vec{A}=(By,0,0)$ our system is
described by the separable Hamiltonian  ($e=|e|$) 
\begin{equation}\label{eq3_3}
  H=\frac{1}{2m}(p_x+eBy)^2+\frac{1}{2m}p_y^2+V(y)\,,
\end{equation}
i.e. allowing to set $\Psi(x,y)=\exp(ikx)\psi(y)$ for the
eigenfunctions. Our ansatz for the whole wavefunction is
\begin{equation}\label{eq3_2}
  \Psi(x,y)=\frac{1}{\sqrt{2\pi}}\exp(ikx)\sum_j a_j(k)\vp(y-jd)\,,
\end{equation}
i.e. we use the ansatz $\ket{\psi(k,n)}=\sum_j a_j(k)\ket{j}$  for
 $\psi(y)$ ($n$ is the
Landau index in the spectrum of (\ref{eq3_1}) or (\ref{eq3_3}) for a 
given $k$) where $\ket{j}$ is
the ground state localized in the $j$--th well of the potential
(corresponds to the Wannier state of the $B=0$ case). We have thus
limited our model just to the lowest band in the modulation direction
by this ansatz. The Fermi level lies always deep below the top of the
modulation potential in our calculations.

Next we use the
{\em tight--binding approximation} (i.e. $\braket{i}{j}=\delta_{i,j}$,
$\bra{i}H\ket{j}=t\delta_{i,j\pm 1}$, $t<0$) and obtain the Hamiltonian
matrix elements (see also Wulf {\em et al.}\cite{wulf:09:1992})
\begin{equation}\label{eq3_1}
  H_{ij}=\frac{\hbar^2}{2m}\cdot K^2\big((k/K)+i\big)^2\delta_{i,j}
         +t\delta_{i,j\pm 1}\,,\quad K=d\frac{eB}{\hbar}\,.
\end{equation}
In our model the real physical system is thus represented by the parameters $t$
(hopping) and $d$ (period) for the structure and of course $B$ for the
magnetic field. 

Note that (up to the scaling of $k$ and energy) the problem
(\ref{eq3_1}) effectively depends only on the single
parameter~$\alpha$:  Eq. \ref{eq3_1} can be written in
dimensionless form 
\begin{eqnarray}\label{eq3_6}
  H_{ij}&=&|t|\left[\alpha^2\big((k/K)+i\big)^2\delta_{i,j}-
         \delta_{i,j\pm  1}\right]\,,\\ \nonumber
         &&\alpha^2=\frac{e^2B^2}{m}\cdot\frac{d^2}{2|t|}=
	 \left(\frac{\hbar\omegaeff}{2|t|}\right)^2\,.
\end{eqnarray}
If we now assume the system to be infinite in the $y$ direction the
matrix problem (\ref{eq3_1}) is mathematically equivalent to the Mathieu
equation 
\begin{equation}\label{eq3_5}
  -\frac{\hbar^2}{2m}\psi''(x)-\psi(x)2|t|\cos Kx=E\psi(x)
\end{equation}
i.e. 1D Schr\"odinger equation with potential
$W(x)=-2|t|\cos{Kx}$. The detailed description of this model can be found in
\cite{davison:1992}.

This allows us to determine the spectrum of (\ref{eq3_1})
qualitatively, and therefore also the DOS, even before we carry out the
numerical calculations.  Let the magnetic field be weak at first
($\alpha\ll 1$, see explanation below in this paragraph).  Bands
$E(k)$ should appear (as a consequence of the Bloch theorem) and we
may limit ourselves to 1MBZ, the first ``magnetic Brillouin zone'' $k\in
(-\frac{1}{2}K, \frac{1}{2}K)$.  For $E\ll 2|t|$ we expect those bands
to be equidistantly spaced and narrow. The former follows from
$W(x)\approx 2|t|(-1+\frac{1}{2}K^2x^2)$ near the potential minimum
(the condition $\alpha\ll 1$ means just that the spacing of such
states is $\ll 2|t|$), the latter is due to the smallness of the
overlap of low--lying states in two neighbouring wells of $W(x)$.
In contrast, for $E\gg 2|t|$ we expect the spectrum to be almost like
the one of 1D free electrons $E(k)=\sum_i \hbar^2(k+iK)^2/2m$.  The
underlying non--constant potential $V$ will manifest itself in the
gaps which open at the Brillouin zone boundaries ($k=\pm\frac{1}{2}K$).

Let us now translate this analysis into terms of the original
problem \ref{eq3_1}. Consider a fixed magnetic field (or constant
$\alpha$). For low energies we obtain nearly equidistant and sharp
Landau levels (free 2D electron gas in magnetic field). At high
energies we get a sum of almost 1D densities of states,
i.e. independent quantum wires (where $B$ plays no role). The
transition occurs around $E=2|t|$. This agrees with the semiclassical theory
except for the narrow gaps in the continuous quasi--free--1D--electron
part of the spectrum. These indicate the magnetic breakdown for
quantizing magnetic fields.

Secondly, we focus on strong magnetic fields ($\alpha\gg 1$).  On one
hand, the states of Eq. \ref{eq3_5} below $2|t|$ will become more
widely spaced and even the lowest state will no longer have $E\ll 2|t|$.
We then expect even the lowest LB to be broad (solid black curve at
Fig. \ref{fig2_1}) because its energy lies in an intermediate region between
$E\ll 2|t|$ and $E\gg 2|t|$. On the other hand the states with $E>2|t|$ which
approach the free electron states (and have a nearly free--electron
parabolic spectrum) will have wider gaps (the smaller the lattice
constant of a crystal, the wider the gaps).

\begin{figure}
\begin{center}

\leavevmode
\hskip5mm
\hbox{\includegraphics[scale=0.4, angle=-90]{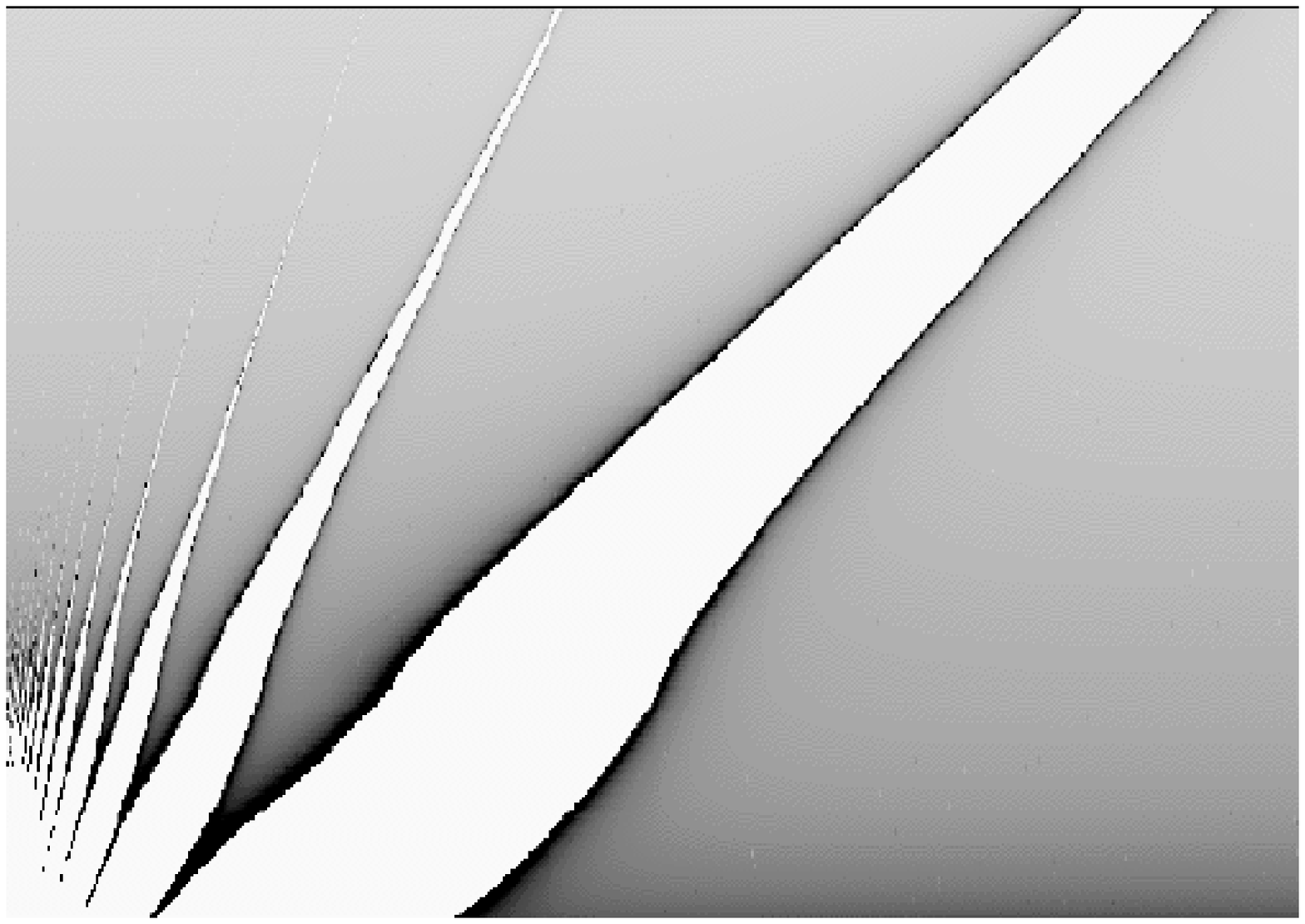}}
\kern-7.54cm\raise.11cm\hbox{\includegraphics[scale=0.327, angle=-90]{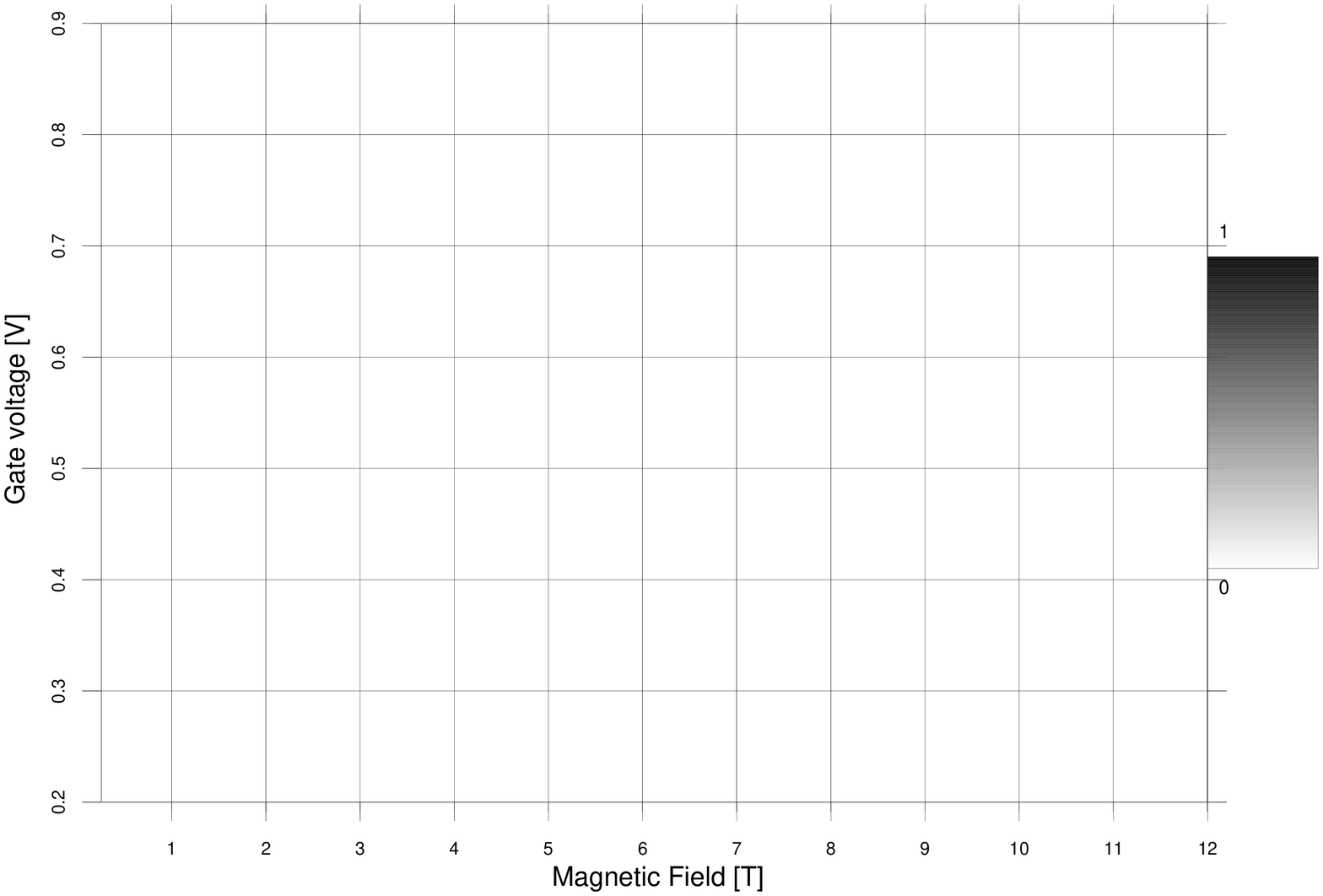}}

\hskip5mm
\hbox{\includegraphics[scale=0.4, angle=-90]{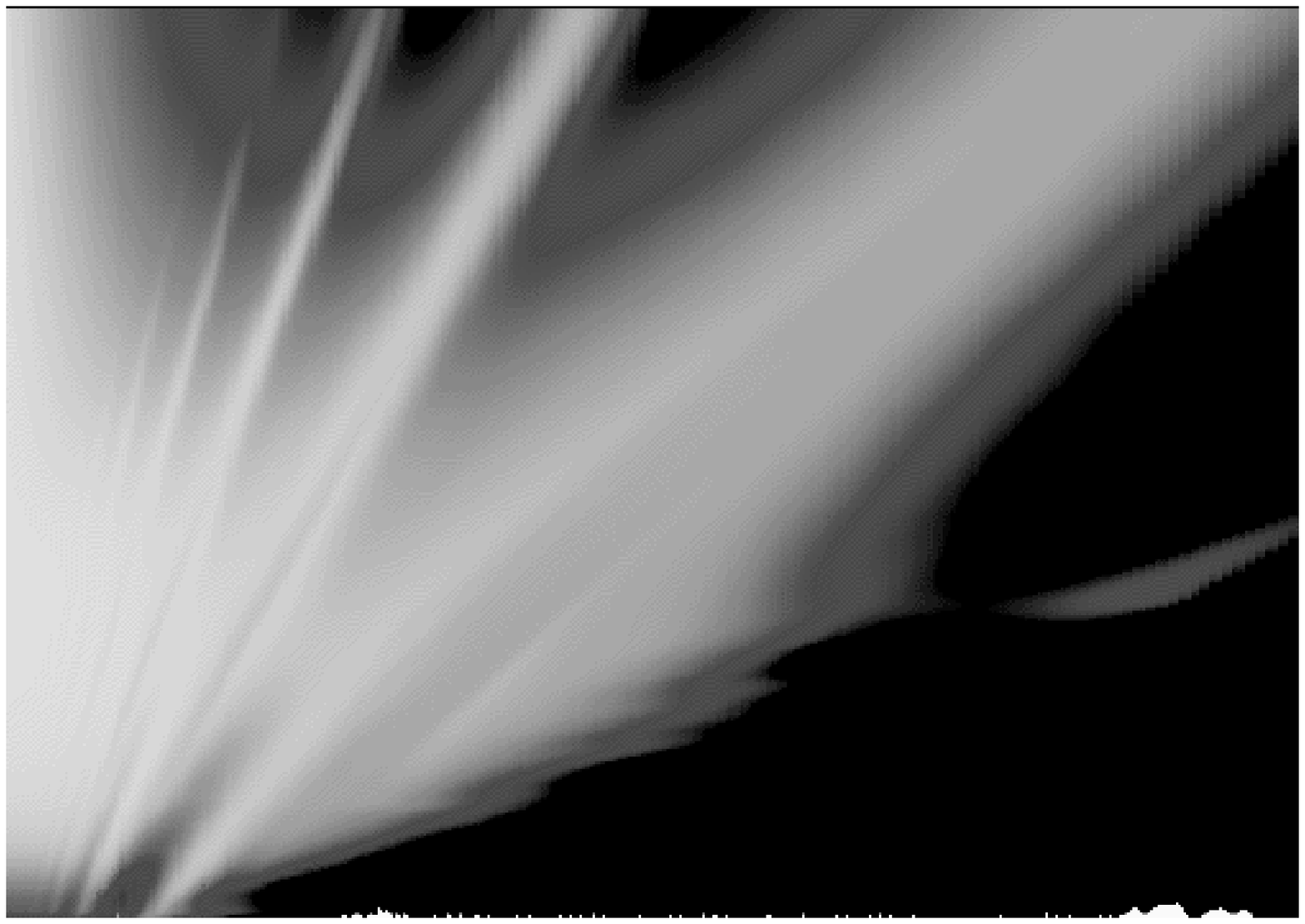}}
\kern-7.54cm\raise.11cm\hbox{\includegraphics[scale=0.327, angle=-90]{fig5.eps}}
\caption{ {\it Above:} density of states (arbitrary units) in the
tight--binding model for different magnetic fields and gate voltages.
{\it Below:} measured resistance (logaritmic scale, arbitrary units).}
\label{fig3_1}
\end{center}
\end{figure}

We may now compare the DOS being output from our model with the
experimentally measured resistance, see Fig. \ref{fig3_1}. In the
experiments, the Fermi energy was adjusted by applying a gate voltage
$U_g$ and   we assume that $E_F$ remains constant while
magnetic field is swept. 

Experiments showed that the gate voltage is proportional to the
concentration of electrons ($N$, see Sec. \ref{secIV}). The
proportionality constant is also approximatelly equal to the capacity
of a parallel--plate capacitor corresponding to the gated
structure. We took the concentration to be equal to the zero--field
concentration and computed the corresponding Fermi energy, 
neglecting the localization effects.  Looking at Fig. \ref{fig3_1}
we see the gaps in the DOS matching very well with the straight lines
of magnetoresistance extrema which justifies this $U_g\leftrightarrow
E_F$ model of ours.

We can see in the Fig. \ref{fig3_1} that varying the magnetic field
and keeping the Fermi energy (or $U_g$) constant and low the system
behaves like an almost free 2D electron gas (sharp LLs, equidistant in
$1/B$) with modified effective mass $\sqrt{mm_y}$ (as predicted by SC
theory). At intermediate $E_F$ however, we observe that the narrow
bands become broad at high $B$ indicating an effective 2D to 1D
transition (when cyclotron radii become smaller than the modulation
period). In contrast, the SC theory states that once the Fermi energy
is set, a magnetic field cannot change the dimensionality of the
system. When $E_F$ is high the quantum wires are decoupled, but there
still open gaps in the continuous spectrum (reflecting the tunneling
between the open SC orbits).

 The lighter region around $B\approx 11\unit{T}$ and $U_g\approx
0.45\unit{V}$ in the experimental data (Fig. \ref{fig3_1} below)
suggests that our model works no longer for strong magnetic fields
($\alpha\gg 1$) and low Fermi energies (even the lowest LB nearly
empty). This condition matches the situation when the cyclotron radius
is much smaller than $d$ and it means that our tight--binding model is
inappropriate for magnetic fields strong enough to create LLs within
one quantum wire (so that even the width of wires will be large enough
for the electrons to behave as a 2DES inside one wire). These are,
however, rather extreme conditions for the experiments shown in
Fig. \ref{fig3_1}.

\section{Transport: Kubo Formula}
\label{secIII}

Knowing the eigenfunctions to the matrix problem \ref{eq3_1} we may
use the linear response theory (Kubo formula, see
e.g. \cite{streda:12:1981}) to compute the conductivity tensor
components (per system area $A$)
\begin{eqnarray*}\label{eq4_1}
  \sigma_{ii}(E_F)&=&\frac{\pi\hbar e^2}{A}
  \Tr\Big[\op{v}_i\delta_\Gamma(E_F-\op{H})\op{v}_i\delta_\Gamma(E_F-\op{H})
\Big]\\
  \sigma_{xy}(E)&=&\ds\frac{i\hbar e^2}{A}\cdot
          	\frac{1}{2}\Tr\Big[\op{v}_x\op{G}^+
          (E)\op{v}_y\delta_\Gamma(E-\op{H})-\\      
&&\hspace{3mm} 
  \op{v}_x\delta_\Gamma(E-\op{H})\op{v}_y\op{G}^-(E)\Big]
  +e\frac{\partial N(E)}{\partial B}\, .
\end{eqnarray*}
Here $\op{v}_{x,y}$ stand for the velocity operator components,
$\delta_\Gamma(E-\op{H})=-(1/2\pi i)\big(\op{G}^+(E)-\op{G}^-(E)\big)$
and $\op{G}^{\pm}(E)=(E-\op{H}\pm i\Gamma)$.  Isotropic 
scattering by a random impurity potential is taken into account by
means of the complex--valued self--energy $\Sigma=\Delta+i\Gamma$ and
we neglect its real part. In the self--consistent Born approximation
and assuming $\Gamma(E,B)$ to be small compared to the level
separation (typically $\hbar\omegaeff$) we can express components
of $\sigma$ by means of the DOS $g(E)$ and matrix elements of $y$
($\omega=eB/m$)
\begin{eqnarray*}
   \sigma_{xx}(E) &=&
  \frac{2}{\pi\Gamma
  d}\cdot\frac{e^2}{h}\cdot\frac{\sgn\big(g(E)\big)}{g(E)}+\\
  \lefteqn{\hskip-8mm\frac{4\pi\Gamma}{d}\cdot
  (\hbar\omega)^2\cdot\frac{e^2}{h}\cdot
  g(E)\sum_{n'\not=n} 
  \left(\frac{\bra{\psi(k,n')} y \ket{\psi(k,n)}}{E(k,n')-
  E(k,n)}\right)^2}\\
  \sigma_{yy}(E)&=&\frac{4\pi \Gamma}{d}\cdot\frac{e^2}{h} 
  g(E)\sum_{n'\not=n} 
  \Big(\bra{\psi(k,n')} y \ket{\psi(k,n)}\Big)^2\\
  \sigma_{xy}(E)&=&e\frac{\partial N(E)}{\partial B}+\\
  \lefteqn{+\frac{4\pi\hbar\omega}{d}\cdot\frac{e^2}{h}g(E)
  \sum_{n'\not= n}
  \left(\bra{\psi(k,n)}y\ket{\psi(k,n')}\right)^2}
\end{eqnarray*}
(including spin degeneracy).  The symbolic expression
$\sgn\big(g(E)\big)/g(E)$ in the first term of $\sigma_{xx}$ indicates
that this term vanishes if $g(E)=0$.  Just to get simpler formulae,
the additional assumption has been made that there are merely two
points at the Fermi level $E=E(k,n)$ within the 1MBZ. We reproduced
the older results of Wulf {\em et al.}\cite{wulf:09:1992} where the
$\Gamma$--independent $\sigma_{xy}(E)$ was calculated within the same
model but in a formally different way (Eqs. 5,6 in
\cite{streda:12:1981}).

 Note the structure of the components of $\sigma$: in general, all
the terms (except for the $\partial N/\partial B$ term in
$\sigma_{xy}$) are products of the DOS ($g$) and some matrix elements
of $\op{y}$. There are two contributions to the conduction parallel to the
wires ($\sigma_{xx}$): the first term (proportional to $1/g$)
originates from the diagonal matrix elements
$\delta(k-k')\hbar k/(eB)+\bra{\psi(k',n)}\op{y}\ket{\psi(k,n)}\propto
\bra{\psi(k',n)}\op{v}_x\ket{\psi(k,n)}=(1/\hbar)(\d E/\d
k)\delta(k-k')\propto 1/g$. It corresponds to the classical
conductivity of a wire (open electron trajectories) or in other words
it does not vanish owing to the unilateral modulation of the system:
without modulation (free 2DES) the DOS comprises of delta--peaks
(sharp LLs) and $(\sgn{} g)/g\to 0$.  For $E<2|t|$, we obtain from our
calculations LBs with non--zero width and this reflects tunnelling
between two closed orbits in the SC picture (see Fig. \ref{fig2_2}),
i.e. indicates a deviation from the quasi--2D behaviour. This
contribution to conductivity is being {\em suppressed} by the impurity
scattering (by $1/\Gamma$) or in other words, persists even if there
is no impurity scattering at all.

The conduction perpendicular to the wires ($\sigma_{yy}$) and
the second term of $\sigma_{xx}$ reflect the inter--LB transitions and
appear thus {\em due to} impurity scattering (they are
proportional to $\Gamma$). These contributions 
might be viewed as a consequence of
tunnelling between the open SC orbits. There is no requirement that
$\bra{\psi(n,k)}y\ket{\psi(m,k)}=0$ for $|n-m|>1$ as in the limit of
weak modulation, but we found these matrix
elements to be decaying rapidly with growing $|n-m|$. In other words,
inter--LB scattering occurs dominantly between neighbouring LBs.

The Hall conductivity does not depend on the scattering in the leading
order at all.

Let us now concentrate on the issue of impurity scattering. Following
\cite{zhang:11:1989,pfannkuche:06:1992,ando:04:1982:p536} we used an
ansatz of self--energy depending only on energy (and $B$). This leads
to a self--consistent equation
$$
  \Sigma(E)\propto\Tr \frac{1}{E-H-\Sigma(E)}
$$
where the proportionality constant describes the strength of the impurity
scattering. This equation yields 
\cite{gerhardts:03:1971,gerhardts:09:1975} the well--known result
$$
  \Gamma^2=\frac{1}{2\pi}\hbar\omega\frac{\hbar}{\tau}
$$
for free 2DES (i.e. no modulation) and short--range scattering
potential. $\tau$ is the relaxation--time in the zero magnetic field
case (as in the Drude theory). In our calculations we used this (i.e.
$\Gamma=\gamma\sqrt{B}$) as a phenomenological ansatz which has
already proven to be useful in explaining the magnetoresistance data
by long--period superlattices \cite{manolescu:08:1995,tornow:12:1996}. 
Surprisingly enough, even this simple ansatz provides a very good
qualitative agreement with the experimental data and makes the results
of the calculations depend on the fitting parameter ($\gamma$,
scattering strength) in a very simple way.

In order to obtain data comparable to experiments, we need to express
the resistivity tensor components
\begin{equation}\label{eq4_3}
  \vr_{yy}=\frac{\sigma_{xx}}{\sigma_{xx}\sigma_{yy}+\sigma_{xy}^2}\,,\quad
  \vr_{xy}=\frac{\sigma_{xy}}{\sigma_{xx}\sigma_{yy}+\sigma_{xy}^2}\,.
\end{equation}
A remarkable and important point of these formulae is that if
scattering is weak (i.e. the inter--LB term $\propto\Gamma$ in
$\sigma_{xx}$ may be neglected), the denominators do not depend on
$\Gamma$.

\begin{figure}
\begin{center}
\includegraphics[angle=-90,scale=0.3]{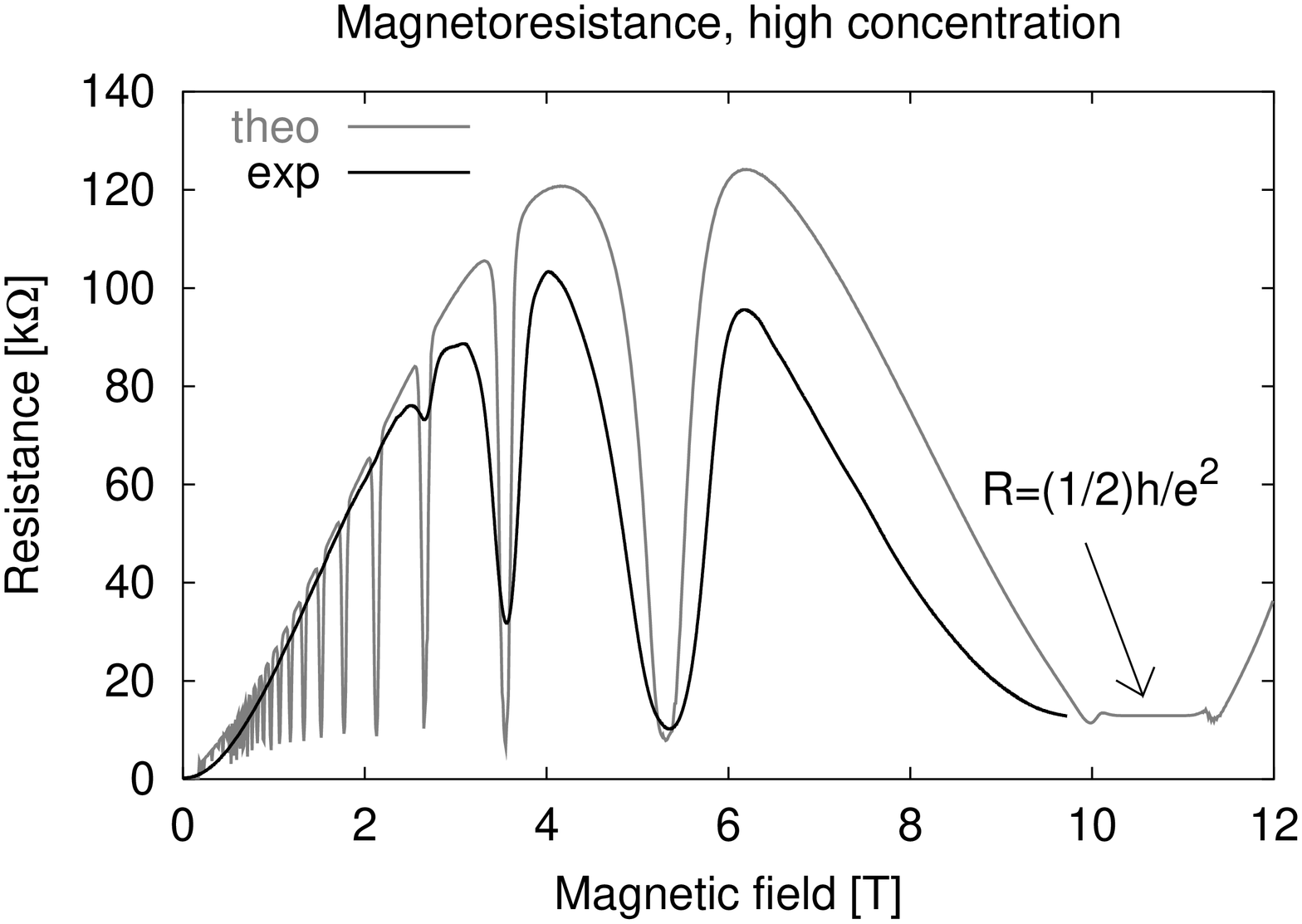}
\medskip
\includegraphics[angle=-90,scale=0.3]{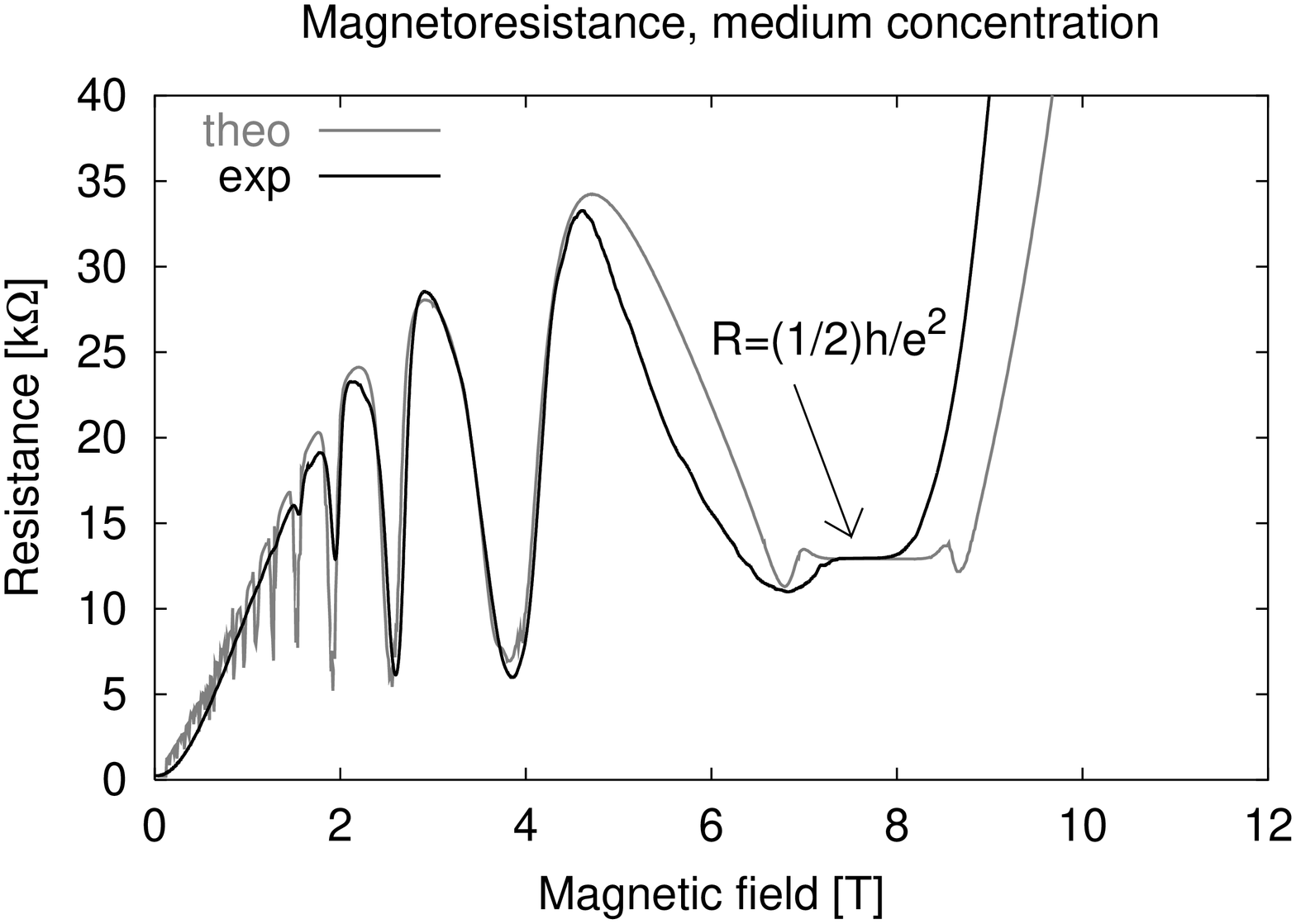}
\medskip
\includegraphics[angle=-90,scale=0.3]{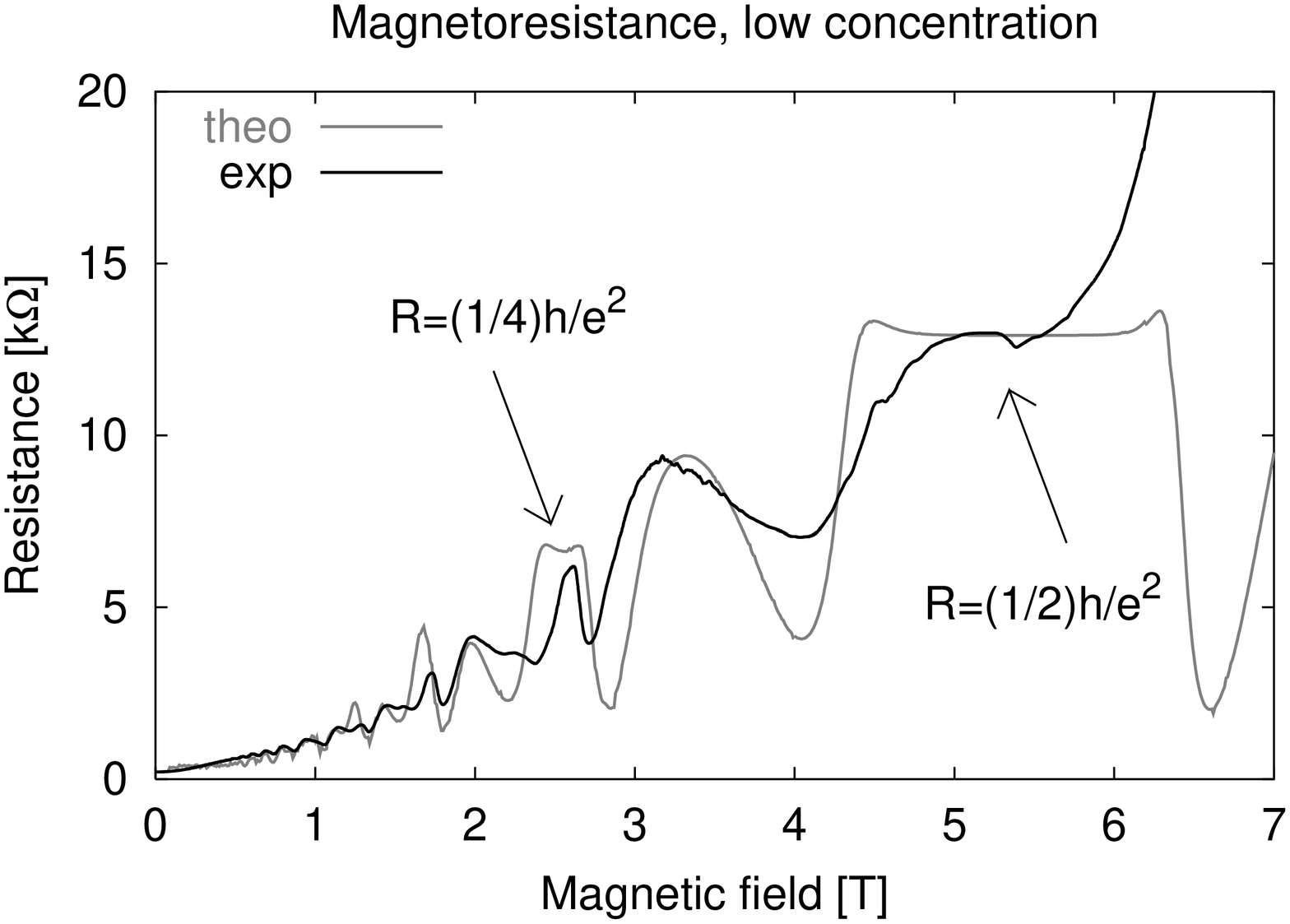}
\end{center}
\caption{Magnetoresistance. Calculations and
experiment. Electron concentrations (from the top)
$5.1\times 10^{11} \unit{cm^{-2}}$, $3.8\times 10^{11}
\unit{cm^{-2}}$, $2.65\times 10^{11} \unit{cm^{-2}}$. The lowest
concentration corresponds to $E_F\approx 2|t|$.}
\label{fig4_1}
\end{figure}

Due to the particular sample geometry a two--point measurement is
performed and thus the experimentally accessible quantity is some
mixture of $\vr_{yy}$ and $\vr_{xy}$. It seems plausible to assume
that the voltage drop measured is a constant linear combination of the
transversal and longitudinal voltage and thus the measured resistance
is
$$
  R=c\vr_{yy}+\vr_{xy}
$$
with some dimensionless constant geometrical factor $c$. Thus if
scattering is weak, the magnetoresistances to be compared to
experimental data contain a single fitting parameter $c/\gamma$. This
comparison for different electron densities is shown at Fig. \ref{fig4_1}.

\section{Discussion}
\label{secIV}

Almost quantitative agreement is achieved for high concentrations
(Fig. \ref{fig4_1}a,b). Exact (no fitting parameter) match between the
magnetoresistance extrema in theory and experiments (which has been
emphasised already on the DOS level) shows that the tight--binding
model captures the essential physics in our experiments. 

Note also that the magnetoresistance approaches quantized values
$R=(1/\nu)h/e^2$ for $E_F$ lying in the gap. First, this allows to
determine the concentration of electrons $N$ in the system (compare
filling factor of each plateau $\nu$ and its $B$). Second, it confirms
our model of the relation between the gate voltage $U_g$ and Fermi
level $E_F$. However, as we may see best in Fig. \ref{fig4_1}c,
there is still room for improving the model e.g. by taking the
localization effects into account.

The quadratic rise of resistivity at low magnetic fields due to
the diagonal component ($\vr_{yy}$) has been predicted within a SC
model \cite{beenakker:04:1989} and measured also at weak--modulation
samples \cite{tornow:12:1996}. Our results (with single fitting
parameter at Fig. \ref{fig4_1}a,b) deviate from this slightly
because of the form of $\Gamma(E,B)\propto\sqrt{B}$.  Assuming
$\Gamma$ independent on $B$ in this region, the quadratic behaviour is
reconciled. For intermediate fields the matrix elements of $\op{y}$ start
to play a role.

Another feature of the tight--binding model is the capability of
explaining why the Hall plateaus show minima at high concentrations and
maxima at low concentrations (Fig. \ref{fig4_1}c). Note, however, that
even at this concentration $E\approx 2|t|$. Swapping the
dominance of $\sigma_{xy}$ and $\sigma_{xx,yy}$ (see esp. Eq. 
\ref{eq4_3}) is the cause for the qualitative change of the
magnetoresistance curve. Nevertheless, we are aware of the feebler
match between theory and experiments at lower concentrations
indicating that our ansatz for the self--energy is rather crude.

\section{Conclusion}

The cleaved--edge--overgrowth technology opens a new area of 2DES with
short--period atomically precise unidirectional modulation for
experimental studies. Such samples may be viewed as an array of
coupled quantum wires which can be decoupled by applying reasonable
($\lesssim 5\unit{T}$) perpendicular magnetic field. From a
theoretical point of view, these systems have a simple band
structure and can be very well described by a simple tight--binding
model. It was demonstrated that for weak magnetic field
($\hbar\omegaeff\ll 2|t|$) the SC approach argumenting with closed and
open electron trajectories is expected to be fairly good and it {\em
is} good indeed. On the other hand, this approach fails for quantizing
magnetic fields (magnetic breakdown in the SC terminology). The fully
quantum--mechanical tight--binding model offers both a good way to
estimate the applicability of the SC approach and a reasonable
description of the system for any magnetic field unless $R_c=k_F\ell^2\ll
d$. This model is also very intuitive on the level of DOS analysis
owing to the analogy with Mathieu equation (\ref{eq3_5}).

We found a good agreement between the experimental magnetoresistance
and calculations based on the linear response theory even with a very
simple model for impurity scattering. We expect that the
match between theory and experiment may be improved by
internaly--consistent treatment of scattering due to 
randomly distributed impurities.

\section{Acknowledgements}

One of the authors (KV) would like to acknowledge discussions with
Daniela Pfannkuche and Alexander Chudnovskiy.  This work has been
partly supported by the Grant Agency of the Czech Republic under
Grants No. 202/01/0754 and No. 202/03/0431.

\bibliography{bib_wires}

\begin{thebibliography}{27}
\expandafter\ifx\csname natexlab\endcsname\relax\def\natexlab#1{#1}\fi
\expandafter\ifx\csname bibnamefont\endcsname\relax
  \def\bibnamefont#1{#1}\fi
\expandafter\ifx\csname bibfnamefont\endcsname\relax
  \def\bibfnamefont#1{#1}\fi
\expandafter\ifx\csname citenamefont\endcsname\relax
  \def\citenamefont#1{#1}\fi
\expandafter\ifx\csname url\endcsname\relax
  \def\url#1{\texttt{#1}}\fi
\expandafter\ifx\csname urlprefix\endcsname\relax\def\urlprefix{URL }\fi
\providecommand{\bibinfo}[2]{#2}
\providecommand{\eprint}[2][]{\url{#2}}

\bibitem[{\citenamefont{Weiss et~al.}(1989)\citenamefont{Weiss, von Klitzing,
  Ploog, and Weinmann}}]{weiss:01:1989}
\bibinfo{author}{\bibfnamefont{D.}~\bibnamefont{Weiss}},
  \bibinfo{author}{\bibfnamefont{K.}~\bibnamefont{von Klitzing}},
  \bibinfo{author}{\bibfnamefont{K.}~\bibnamefont{Ploog}}, \bibnamefont{and}
  \bibinfo{author}{\bibfnamefont{G.}~\bibnamefont{Weinmann}},
  \bibinfo{journal}{Europhys. Lett.} \textbf{\bibinfo{volume}{8}},
  \bibinfo{pages}{179} (\bibinfo{year}{1989}).

\bibitem[{\citenamefont{Beenakker}(1989)}]{beenakker:04:1989}
\bibinfo{author}{\bibfnamefont{C.}~\bibnamefont{Beenakker}},
  \bibinfo{journal}{Phys. Rev. Lett.} \textbf{\bibinfo{volume}{62}},
  \bibinfo{pages}{2020} (\bibinfo{year}{1989}).

\bibitem[{\citenamefont{{St\v reda} and MacDonald}(1990)}]{streda:06:1990}
\bibinfo{author}{\bibfnamefont{P.}~\bibnamefont{{St\v reda}}} \bibnamefont{and}
  \bibinfo{author}{\bibfnamefont{A.}~\bibnamefont{MacDonald}},
  \bibinfo{journal}{Phys. Rev. B.}
  \textbf{\bibinfo{volume}{41}}(\bibinfo{number}{17}), \bibinfo{pages}{11892}
  (\bibinfo{year}{1990}).

\bibitem[{\citenamefont{Gerhardts et~al.}(1989)\citenamefont{Gerhardts, Weiss,
  and v.~Klitzing}}]{gerhardts:03:1989}
\bibinfo{author}{\bibfnamefont{R.}~\bibnamefont{Gerhardts}},
  \bibinfo{author}{\bibfnamefont{D.}~\bibnamefont{Weiss}}, \bibnamefont{and}
  \bibinfo{author}{\bibfnamefont{K.}~\bibnamefont{v.~Klitzing}},
  \bibinfo{journal}{Phys. Rev. Lett.}
  \textbf{\bibinfo{volume}{62}}(\bibinfo{number}{10}), \bibinfo{pages}{1173}
  (\bibinfo{year}{1989}).

\bibitem[{\citenamefont{Vasilopoulos and Peeters}(1989)}]{vasilo:11:1989}
\bibinfo{author}{\bibfnamefont{P.}~\bibnamefont{Vasilopoulos}}
  \bibnamefont{and} \bibinfo{author}{\bibfnamefont{F.}~\bibnamefont{Peeters}},
  \bibinfo{journal}{Phys. Rev. Lett.}
  \textbf{\bibinfo{volume}{63}}(\bibinfo{number}{19}), \bibinfo{pages}{2120}
  (\bibinfo{year}{1989}).

\bibitem[{\citenamefont{Beton et~al.}(1990)\citenamefont{Beton, Alves, Main,
  Eaves, Dellow, Henini, and Hughes}}]{beton:11:1990}
\bibinfo{author}{\bibfnamefont{P.}~\bibnamefont{Beton}},
  \bibinfo{author}{\bibfnamefont{E.}~\bibnamefont{Alves}},
  \bibinfo{author}{\bibfnamefont{P.}~\bibnamefont{Main}},
  \bibinfo{author}{\bibfnamefont{L.}~\bibnamefont{Eaves}},
  \bibinfo{author}{\bibfnamefont{M.}~\bibnamefont{Dellow}},
  \bibinfo{author}{\bibfnamefont{M.}~\bibnamefont{Henini}}, \bibnamefont{and}
  \bibinfo{author}{\bibfnamefont{O.}~\bibnamefont{Hughes}},
  \bibinfo{journal}{Phys. Rev. B.}
  \textbf{\bibinfo{volume}{42}}(\bibinfo{number}{14}), \bibinfo{pages}{9229}
  (\bibinfo{year}{1990}).

\bibitem[{\citenamefont{Shi and Szeto}(1996)}]{shi:05:1996}
\bibinfo{author}{\bibfnamefont{Q.}~\bibnamefont{Shi}} \bibnamefont{and}
  \bibinfo{author}{\bibfnamefont{K.}~\bibnamefont{Szeto}},
  \bibinfo{journal}{Phys. Rev. B.}
  \textbf{\bibinfo{volume}{53}}(\bibinfo{number}{19}), \bibinfo{pages}{12990}
  (\bibinfo{year}{1996}).

\bibitem[{\citenamefont{Menne and Gerhardts}(1998)}]{menne:01:1998}
\bibinfo{author}{\bibfnamefont{R.}~\bibnamefont{Menne}} \bibnamefont{and}
  \bibinfo{author}{\bibfnamefont{R.}~\bibnamefont{Gerhardts}},
  \bibinfo{journal}{Phys. Rev. B.}
  \textbf{\bibinfo{volume}{57}}(\bibinfo{number}{3}), \bibinfo{pages}{1707}
  (\bibinfo{year}{1998}).

\bibitem[{\citenamefont{Zwerschke and Gerhardts}(1998)}]{zwerschke:00:1998}
\bibinfo{author}{\bibfnamefont{S.}~\bibnamefont{Zwerschke}} \bibnamefont{and}
  \bibinfo{author}{\bibfnamefont{R.}~\bibnamefont{Gerhardts}},
  \bibinfo{journal}{Physica E} (\bibinfo{number}{256-258}), \bibinfo{pages}{28}
  (\bibinfo{year}{1998}).

\bibitem[{\citenamefont{Zwerschke et~al.}(1998)\citenamefont{Zwerschke,
  Manolescu, and Gerhardts}}]{zwerschke:08:1998}
\bibinfo{author}{\bibfnamefont{S.}~\bibnamefont{Zwerschke}},
  \bibinfo{author}{\bibfnamefont{A.}~\bibnamefont{Manolescu}},
  \bibnamefont{and}
  \bibinfo{author}{\bibfnamefont{R.}~\bibnamefont{Gerhardts}},
  \bibinfo{journal}{Phys. Rev. B.}
  \textbf{\bibinfo{volume}{60}}(\bibinfo{number}{8}), \bibinfo{pages}{5536}
  (\bibinfo{year}{1998}).

\bibitem[{\citenamefont{{M\"uller}}(1992)}]{muller:01:1992}
\bibinfo{author}{\bibfnamefont{J.}~\bibnamefont{{M\"uller}}},
  \bibinfo{journal}{Phys. Rev. Lett.}
  \textbf{\bibinfo{volume}{68}}(\bibinfo{number}{3}), \bibinfo{pages}{386}
  (\bibinfo{year}{1992}).

\bibitem[{\citenamefont{Peeters and Vasilopoulos}(1993)}]{peeters:01:1993}
\bibinfo{author}{\bibfnamefont{F.}~\bibnamefont{Peeters}} \bibnamefont{and}
  \bibinfo{author}{\bibfnamefont{P.}~\bibnamefont{Vasilopoulos}},
  \bibinfo{journal}{Phys. Rev. B.}
  \textbf{\bibinfo{volume}{47}}(\bibinfo{number}{3}), \bibinfo{pages}{1466}
  (\bibinfo{year}{1993}).

\bibitem[{\citenamefont{Deutschmann et~al.}(2000)\citenamefont{Deutschmann,
  Lorke, Wegscheider, Bichler, and Abstreiter}}]{deutschmann:02:2000}
\bibinfo{author}{\bibfnamefont{R.~A.} \bibnamefont{Deutschmann}},
  \bibinfo{author}{\bibfnamefont{A.}~\bibnamefont{Lorke}},
  \bibinfo{author}{\bibfnamefont{W.}~\bibnamefont{Wegscheider}},
  \bibinfo{author}{\bibfnamefont{M.}~\bibnamefont{Bichler}}, \bibnamefont{and}
  \bibinfo{author}{\bibfnamefont{G.}~\bibnamefont{Abstreiter}},
  \bibinfo{journal}{Physica E}
  \textbf{\bibinfo{volume}{6}}(\bibinfo{number}{1-4}), \bibinfo{pages}{561}
  (\bibinfo{year}{2000}).

\bibitem[{\citenamefont{Deutschmann et~al.}(2001)\citenamefont{Deutschmann,
  Wegscheider, Rother, Bichler, Abstreiter, Albrecht, and
  Smet}}]{deutschmann:02:2001}
\bibinfo{author}{\bibfnamefont{R.~A.} \bibnamefont{Deutschmann}},
  \bibinfo{author}{\bibfnamefont{W.}~\bibnamefont{Wegscheider}},
  \bibinfo{author}{\bibfnamefont{M.}~\bibnamefont{Rother}},
  \bibinfo{author}{\bibfnamefont{M.}~\bibnamefont{Bichler}},
  \bibinfo{author}{\bibfnamefont{G.}~\bibnamefont{Abstreiter}},
  \bibinfo{author}{\bibfnamefont{C.}~\bibnamefont{Albrecht}}, \bibnamefont{and}
  \bibinfo{author}{\bibfnamefont{J.~H.} \bibnamefont{Smet}},
  \bibinfo{journal}{Phys. Rev. Lett.}
  \textbf{\bibinfo{volume}{86}}(\bibinfo{number}{9}), \bibinfo{pages}{1857}
  (\bibinfo{year}{2001}).

\bibitem[{\citenamefont{Deutschmann}(2001)}]{deutschmann:2001}
\bibinfo{author}{\bibfnamefont{R.~A.} \bibnamefont{Deutschmann}},
  \emph{\bibinfo{title}{Two dimensional electron systems in atomically precise
  periodic potentials}}, Selected topics of Semiconductor Physics and
  Technology, {Vol. 42} (\bibinfo{publisher}{University of Technology Munich},
  \bibinfo{year}{2001}), \bibinfo{note}{{PhD. thesis, ISBN 3-932749-42-1}}.

\bibitem[{\citenamefont{Ashcroft and Mermin}(1976)}]{ashcroft:1976:p232}
\bibinfo{author}{\bibfnamefont{N.}~\bibnamefont{Ashcroft}} \bibnamefont{and}
  \bibinfo{author}{\bibfnamefont{N.}~\bibnamefont{Mermin}},
  \emph{\bibinfo{title}{Solid State Physics}} (\bibinfo{publisher}{Saunders
  College Publishing}, \bibinfo{address}{Orlando}, \bibinfo{year}{1976}),
  \bibinfo{note}{p.~232}.

\bibitem[{\citenamefont{Gradshteyn and Ryzhik}(1980)}]{ryzhik:1980:p904}
\bibinfo{author}{\bibfnamefont{I.}~\bibnamefont{Gradshteyn}} \bibnamefont{and}
  \bibinfo{author}{\bibfnamefont{I.}~\bibnamefont{Ryzhik}},
  \emph{\bibinfo{title}{Table of Integrals, Series and Products}}
  (\bibinfo{publisher}{Academic Press}, \bibinfo{address}{San Diego},
  \bibinfo{year}{1980}), \bibinfo{note}{p.~904}.

\bibitem[{\citenamefont{Pfannkuche and Gerhardts}(1992)}]{pfannkuche:06:1992}
\bibinfo{author}{\bibfnamefont{D.}~\bibnamefont{Pfannkuche}} \bibnamefont{and}
  \bibinfo{author}{\bibfnamefont{R.}~\bibnamefont{Gerhardts}},
  \bibinfo{journal}{Phys. Rev. B.}
  \textbf{\bibinfo{volume}{46}}(\bibinfo{number}{19}), \bibinfo{pages}{12606}
  (\bibinfo{year}{1992}).

\bibitem[{\citenamefont{Wulf et~al.}(1992)\citenamefont{Wulf, {J. Ku\v cera},
  and MacDonald}}]{wulf:09:1992}
\bibinfo{author}{\bibfnamefont{U.}~\bibnamefont{Wulf}},
  \bibinfo{author}{\bibnamefont{{J. Ku\v cera}}}, \bibnamefont{and}
  \bibinfo{author}{\bibfnamefont{A.}~\bibnamefont{MacDonald}},
  \bibinfo{journal}{Phys. Rev. B.}
  \textbf{\bibinfo{volume}{47}}(\bibinfo{number}{3}), \bibinfo{pages}{1675}
  (\bibinfo{year}{1992}).

\bibitem[{\citenamefont{Davidson and St\'{e}\c{s}licka}(1992)}]{davison:1992}
\bibinfo{author}{\bibfnamefont{S.}~\bibnamefont{Davidson}} \bibnamefont{and}
  \bibinfo{author}{\bibfnamefont{M.}~\bibnamefont{St\'{e}\c{s}licka}},
  \emph{\bibinfo{title}{Basic Theory of Surface States}}
  (\bibinfo{publisher}{Claredon Press}, \bibinfo{address}{Oxford},
  \bibinfo{year}{1992}), \bibinfo{note}{p.~41}.

\bibitem[{\citenamefont{{St\v reda}}(1982)}]{streda:12:1981}
\bibinfo{author}{\bibfnamefont{P.}~\bibnamefont{{St\v reda}}},
  \bibinfo{journal}{J. Phys. C: Solid State Phys.}
  \textbf{\bibinfo{volume}{15}}, \bibinfo{pages}{L717} (\bibinfo{year}{1982}).

\bibitem[{\citenamefont{Zhang and Gerhardts}(1990)}]{zhang:11:1989}
\bibinfo{author}{\bibfnamefont{C.}~\bibnamefont{Zhang}} \bibnamefont{and}
  \bibinfo{author}{\bibfnamefont{R.}~\bibnamefont{Gerhardts}},
  \bibinfo{journal}{Phys. Rev. B.}
  \textbf{\bibinfo{volume}{41}}(\bibinfo{number}{18}), \bibinfo{pages}{12850}
  (\bibinfo{year}{1990}).

\bibitem[{\citenamefont{Ando et~al.}(1982)\citenamefont{Ando, Fowler, and
  Stern}}]{ando:04:1982:p536}
\bibinfo{author}{\bibfnamefont{T.}~\bibnamefont{Ando}},
  \bibinfo{author}{\bibfnamefont{A.}~\bibnamefont{Fowler}}, \bibnamefont{and}
  \bibinfo{author}{\bibfnamefont{F.}~\bibnamefont{Stern}},
  \bibinfo{journal}{Rev. Mod. Phys.} \textbf{\bibinfo{volume}{54}},
  \bibinfo{pages}{536} (\bibinfo{year}{1982}).

\bibitem[{\citenamefont{Gerhardts and Hajdu}(1971)}]{gerhardts:03:1971}
\bibinfo{author}{\bibfnamefont{R.}~\bibnamefont{Gerhardts}} \bibnamefont{and}
  \bibinfo{author}{\bibfnamefont{J.}~\bibnamefont{Hajdu}},
  \bibinfo{journal}{Zeitschrift {f\" ur} Physik}
  \textbf{\bibinfo{volume}{245}}, \bibinfo{pages}{126} (\bibinfo{year}{1971}).

\bibitem[{\citenamefont{Gerhardts}(1975)}]{gerhardts:09:1975}
\bibinfo{author}{\bibfnamefont{R.}~\bibnamefont{Gerhardts}},
  \bibinfo{journal}{Zeitschrift {f\" ur} Physik B}
  \textbf{\bibinfo{volume}{22}}, \bibinfo{pages}{327} (\bibinfo{year}{1975}).

\bibitem[{\citenamefont{Manolescu et~al.}(1996)\citenamefont{Manolescu,
  Gerhardts, Tornow, Weiss, von Klitzing, and Weinmann}}]{manolescu:08:1995}
\bibinfo{author}{\bibfnamefont{A.}~\bibnamefont{Manolescu}},
  \bibinfo{author}{\bibfnamefont{R.}~\bibnamefont{Gerhardts}},
  \bibinfo{author}{\bibfnamefont{M.}~\bibnamefont{Tornow}},
  \bibinfo{author}{\bibfnamefont{D.}~\bibnamefont{Weiss}},
  \bibinfo{author}{\bibfnamefont{K.}~\bibnamefont{von Klitzing}},
  \bibnamefont{and} \bibinfo{author}{\bibfnamefont{G.}~\bibnamefont{Weinmann}},
  \bibinfo{journal}{Surf. Sci.} \textbf{\bibinfo{volume}{361/362}},
  \bibinfo{pages}{513} (\bibinfo{year}{1996}).

\bibitem[{\citenamefont{Tornow et~al.}(1996)\citenamefont{Tornow, Weiss,
  Manolescu, Menne, and von Klitzing}}]{tornow:12:1996}
\bibinfo{author}{\bibfnamefont{M.}~\bibnamefont{Tornow}},
  \bibinfo{author}{\bibfnamefont{D.}~\bibnamefont{Weiss}},
  \bibinfo{author}{\bibfnamefont{A.}~\bibnamefont{Manolescu}},
  \bibinfo{author}{\bibfnamefont{R.}~\bibnamefont{Menne}}, \bibnamefont{and}
  \bibinfo{author}{\bibfnamefont{K.}~\bibnamefont{von Klitzing}},
  \bibinfo{journal}{Phys. Rev. B.}
  \textbf{\bibinfo{volume}{54}}(\bibinfo{number}{23}), \bibinfo{pages}{16397}
  (\bibinfo{year}{1996}).

\end{thebibliography}

\end{document}